\documentclass[9pt]{article}
\usepackage{float}
\usepackage[symbol]{footmisc}
\usepackage[superscript,sort]{cite}
\usepackage{array}
\usepackage[pdftex,colorlinks=true,linkcolor=black,citecolor=black,urlcolor=black,bookmarks=true,breaklinks=true]{hyperref}
\usepackage[pdftex]{graphicx}
\usepackage[usenames,pdftex]{color}
\definecolor{Red}{rgb}{1.0,0.0,0.0}
\usepackage{amsmath,amssymb}

\usepackage{amsthm,amscd,amsxtra,amsfonts,amsmath,amssymb,multirow}
\usepackage{wrapfig}
\usepackage[footnotesize]{caption}
\usepackage[tiny,compact]{titlesec}
\usepackage{ctable}
\usepackage{xcolor,colortbl}
\usepackage{subfigure}
\usepackage{tikz}
\usetikzlibrary{shapes,arrows}
\usepackage[T1]{fontenc}

\titlespacing*{\section}{0pt}{*0}{*0}
\titlespacing*{\subsection}{0pt}{*0}{*0}
\titlespacing*{\subsubsection}{0pt}{*0}{*0}
\titlespacing{\paragraph}{0pt}{*0}{*1}

\definecolor{MyPurple}{rgb}{1,0,1}

\newcommand{\beq}[1]{\begin{equation} \label{#1}}
\newcommand{\eeq}{\end{equation}}
\newcommand{\barray}{\begin{array}{ll}}
\newcommand{\earray}{\end{array}}

\begin{document}
\pagenumbering{roman}

 %\clearpage \pagebreak \setcounter{page}{1}
 \renewcommand{\thepage}{{\arabic{page}}}

\title{Correlation function based   Gaussian network models
}

\author{
Kelin Xia$^1$, Kristopher Opron$^2$ and
Guo-Wei Wei$^{1,2,3}$ \footnote{ Address correspondences  to Guo-Wei Wei. E-mail:wei@math.msu.edu}\\
%\address{
$^1$ Department of Mathematics \\
Michigan State University, MI 48824, USA\\
$^2$  Department of Biochemistry and Molecular Biology\\
Michigan State University, MI 48824, USA \\
$^3$ Department of Electrical and Computer Engineering \\
Michigan State University, MI 48824, USA \\
}

\date{\today}
\maketitle

\begin{abstract}

Gaussian network model (GNM) is one of the most accurate and efficient methods for biomolecular flexibility analysis. However, the systematic generalization of the GNM has been elusive. We show that the GNM Kirchhoff matrix can be built  from the ideal low-pass filter, which is a special case of a wide class of correlation functions underpinning the linear scaling flexibility-rigidity index (FRI) method. Based on the mathematical structure of  correlation functions,  we propose a unified framework to construct generalized Kirchhoff matrices whose matrix inverse leads to correlation function based GNMs, whereas, the direct inverse of the diagonal elements gives rise to FRI method. We illustrate that  correlation function based GNMs outperform the original GNM in the B-factor prediction of a set of 364 proteins. We demonstrate that for any given correlation function, FRI and GNM methods provide essentially identical B-factor predictions when the scale value in the correlation function is sufficiently large. 

% However, the FRI method is thousands of times more efficient than the GNM method for the above test set.   

\end{abstract}
Key words:
Gaussian network model,
Flexibility-rigidity index,
Thermal fluctuation
%PACS numbers: 87.10.-e, 87.14.et,  87.15.Ya, 02.10.Yn

%\section{Introduction}\label{sec:Intro}

\twocolumn

Under physiological condition, proteins undergo everlasting motions, ranging from atomic thermal fluctuation, side-chain rotation, residue swiveling, to domain swirling. protein motion strongly correlates with protein functions, including  molecular docking \cite{Fischer:2014}, drug binding \cite{Alvarez-Garcia:2014},  allosteric signaling \cite{ZBu:2011}, self assembly \cite{Marsh:2014} and enzyme catalysis \cite{Fraser:2009}. 
The range of protein motions in a cellular environment depends on the structure's local flexibility, an intrinsic property of a given protein structure. Protein flexibility is reflected by the     Debye-Waller or B-factor, i.e., the atomic mean-square displacement, obtained in  structure determination by x-ray crystallography, NMR, or single-molecule force experiments \cite{Dudko:2006}. However, the B-factor cannot absolutely quantify flexibility: it also depends the crystal environment, solvent type, data collection condition and structural refinement procedure \cite{Kondrashov:2007,Hinsen:2008}.

%Theoretically, biomolecular flexibility can be  investigated by a number of   approaches.   Molecular dynamics (MD) \cite{McCammon:1977}  elucidates biomolecular collective motion and fluctuation.
The flexibility of a biomolecule can be  assessed by  normal mode analysis (NMA)  \cite{Go:1983,Tasumi:1982,Brooks:1983,Levitt:1985}, elastic network model (ENM) \cite{Tirion:1996,Hinsen:1998}, Gaussian network model (GNM) \cite{Bahar:1997,Bahar:1998}, anisotropic network model (ANM) \cite{Atilgan:2001},    graph theory \cite{Jacobs:2001}, etc. NMA  can be regarded as time-independent molecular dynamics (MD) \cite{JKPark:2013}. NMA diagonalizes the MD potential to obtain a set of eigenvalues and eigenvectors, where first  few  eigenvectors predict the collective, global motions,   which are  potentially relevant to   biomolecular functionality. NMA with only the elasticity potential, which is a leading term in the MD potential, was introduced by Tirion \cite{Tirion:1996}, and was extended to the network setting in ANM \cite{Atilgan:2001}.  Here network refers to the connectivity between particles regardless of their chemical bonds \cite{Flory:1976}. The GNM is a highly accurate network based flexibility method. Although it was originally advocated as an ENM \cite{Bahar:1997} and interpreted with the random Gaussian network theory \cite{Flory:1976}, the GNM is strictly not an elastic model \cite{Thorpe:2007}--- it utilizes a Kirchhoff matrix, rather than the harmonic potential of elasticity. Additionally, its computational procedure does not directly invoke  the random Gaussian network theory. Due to the lack of in-depth understanding, there is no rigorous analysis and/or  systematic  generalization of the GNM in the literature. Therefore, there is a pressing need to better understand the working principle of the GNM theory, which is crucial for its further improvement.

A common feature of the aforementioned approaches is that, they all depend on the mode decomposition of the potential matrix, which typically has the computational complexity of $O(N^3)$, where $N$ is the number of elements in the potential matrix. Yang et al. \cite{LWYang:2008} demonstrated that due to its network setting, the GNM is about one order more efficient than most other flexibility approaches.

Recently, we have proposed a few mode-decomposition free methods for flexibility  analysis, i.e.,  molecular nonlinear dynamics \cite{KLXia:2013b},  stochastic dynamics \cite{KLXia:2013f} and flexibility-rigidity index  (FRI)  \cite{KLXia:2013d,Opron:2014}. Among them, the FRI is of $O(N^2)$ in computational complexity and has been accelerated to $O(N)$ without  loss of accuracy  \cite{Opron:2014}. The essential idea of the FRI method is to evaluate the rigidity index or the compactness of the biomolecular (network) packing  by the total correlation. Then the flexibility index is defined as the inverse of the rigidity index. The correlation between any two atoms or residues is measured through correlation functions.  The FRI can be regarded as a generalization of Halle's local density model \cite{Halle:2002}. The FRI method has been shown to be orders of magnitude more efficient and about ten percent more accurate than the GNM for the B-factor prediction of  a set of 365 proteins \cite{Opron:2014}.

The objective of the present work is to shed light on the GNM and FRI methods. Specifically,  we reveal that the ideal low filter used in the GNM Kirchhoff matrix is a special admissible FRI correlation  function, which is the limiting case of many commonly used FRI correction functions. This finding  paves the way for understanding the connection between the GNM and FRI methods.  Additionally, we introduce a generalized Kirchhoff matrix to provide a unified starting point for the GNM and FRI methods, which  throws light on the similarity and difference between GNM and FRI. Moreover, based on this new  understanding of the GNM working principle, we   propose infinitely many  correlation function based GNM methods.  Finally,  we unveil that the FRI and the GNM are asymptotically equivalent when the cutoff value in the Kirchhoff matrix or the scale value in the correlation function is sufficiently large.  The present work offers a new strategy for the design and construction of  accurate, efficient and robust  methods for biomolecular flexibility analysis.

To establish notation and facilitate new development, let us present a brief review of the GNM and FRI methods. Consider an  $N$-particle coarse-grained representation of a biomolecule. We denote $\{ {\bf r}_{i}| {\bf r}_{i}\in \mathbb{R}^{3}, i=1,2,\cdots, N\}$ the coordinates of these particles and $r_{ij}=\|{\bf r}_i-{\bf r}_j\|$ the Euclidean space distance between  $i$th   and $j$th particles. In a nutshell, the GNM prediction of the $i$th B-factor of the biomolecule can be expressed as \cite{Bahar:1997,Bahar:1998}
\begin{eqnarray}\label{eqn:GNM}
B_i^{\rm GNM}=a \left(\Gamma^{-1} \right)_{ii}, \forall i=1,2,\cdots, N,
\end{eqnarray}
where $a$ is a fitting parameter that can be related to  the thermal energy and $\left(\Gamma^{-1} \right)_{ii}$ is the $i$th diagonal element of the matrix inverse of the Kirchhoff matrix,
\begin{eqnarray}\label{eqn:Kirchhoff}
\Gamma_{ij}  = \begin{cases}\begin{array}{ll}
       -1, &i\neq j ~{\rm and} ~r_{ij} \leq r_c \\
       0,  &    i\neq j ~{\rm and} ~r_{ij}  > r_c  \\
        -\sum_{j, j\neq i}^N\Gamma_{ij},  & i=j
							\end{array}
       \end{cases},
\end{eqnarray}
where $r_c$ is a cutoff distance.  The GNM theory evaluates the  matrix inverse by  $\left(\Gamma^{-1} \right)_{ii}=\sum_{k=2}^N  \lambda^{-1}\left[{\bf u}_k {\bf u}_k^T \right]_{ii}$, where $T$ is the transpose and $\lambda_k$ and ${\bf u}_k$ are the $k$th eigenvalue and eigenvector of $\Gamma$, respectively. The summation omits the first eignmode whose eigenvalue is zero.

The FRI   prediction of the $i$th B-factor of the biomolecule can be given by \cite{KLXia:2013d,Opron:2014}
\begin{eqnarray}\label{eqn:FRI}
B_i^{\rm FRI}=a \frac{1}{\sum_{j,j\neq i}^N w_j\Phi(r_{ij};\eta)} + b, \forall i=1,2,\cdots, N,
\end{eqnarray}
where $a$ and $b$ are fitting parameters,   $f_i=\frac{1}{\sum_{j,j\neq i}^N w_j\Phi(r_{ij};\eta)}$ is the $i$th flexibility   index and $\mu_i=\sum_{j,j\neq i}^N w_j\Phi(r_{ij};\eta)$ is the $i$th rigidity index.
Here, $w_j$ is an atomic number depended weight function that can be set to $w_j=1$ for a C$_{\alpha}$ network, and $\Phi(r_{ij};\eta)$ is  a    real-valued monotonically decreasing correlation function satisfying the following admissibility conditions
\begin{eqnarray}\label{eq:couple_matrix1-1}
\Phi( r_{ij};\eta)&=&1 \quad {\rm as }\quad  r_{ij}   \rightarrow 0\\ \label{eq:couple_matrix1-2}
\Phi( r_{ij};\eta)&=&0 \quad {\rm as }\quad  r_{ij}   \rightarrow\infty,
\end{eqnarray}
where $\eta$ is a scale parameter. Delta sequences of the positive type  \cite{GWei:2000} are good choices. Many radial basis functions are also admissible   \cite{KLXia:2013d,Opron:2014}.  Commonly used FRI correlation functions include the  generalized exponential  functions
\begin{eqnarray}\label{eq:couple_matrix1}
\Phi( r_{ij};\eta, \kappa) =    e^{-\left( r_{ij} /\eta \right)^\kappa},    \quad \kappa >0
\end{eqnarray}
and  generalized Lorentz functions
\begin{eqnarray}\label{eq:couple_matrix2}
 \Phi( r_{ij};\eta, \upsilon) =  \frac{1}{1+ \left( r_{ij} /\eta\right)^{\upsilon}},  \quad  \upsilon >0.
 \end{eqnarray}
A major advantage of the FRI method is that it does not resort to  mode decomposition and its computational complexity can be reduced to $O(N)$ by means of the cell lists algorithm used in our fast FRI (fFRI) \cite{Opron:2014}. In contrast, the mode decomposition of NMA and GNM has the computational complexity of $O(N^3)$.

 \begin{figure}[]
\begin{center}
\begin{tabular}{c}
\includegraphics[width=0.49\textwidth]{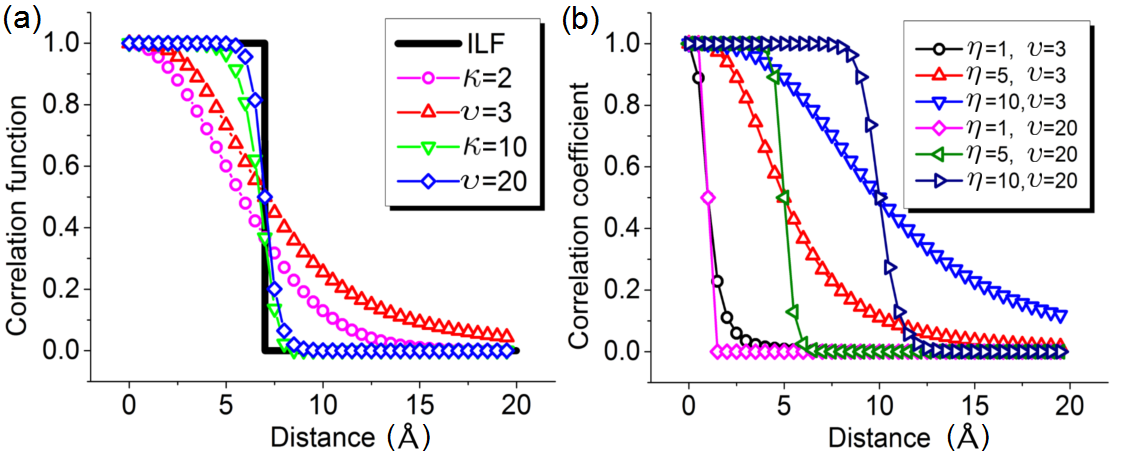}
\end{tabular}
\end{center}
\caption{Illustration of admissible correlation functions.
(a) Correlation  functions approach the ILF as $\kappa\rightarrow \infty$ or  $\upsilon\rightarrow \infty$ at $\eta=7$\AA.
(b) Effects of varying scale value $\eta$. Local correlation is obtained with large $\upsilon$ and small $\eta$ values. Whereas, nonlocal correlation is generated by small $\upsilon$ and large $\eta$ values.  
 }
\label{functions}
\end{figure}
To further explore the theoretical foundation of GNM, let us examine the parameter limits of   generalized exponential functions (\ref{eq:couple_matrix1}) and generalized Lorentz functions (\ref{eq:couple_matrix2})
\begin{eqnarray}\label{eq:Asmpt1}
e^{-\left( r_{ij} /\eta\right)^\kappa} \rightarrow \Phi( r_{ij};r_c)  & {\rm as } & \kappa\rightarrow\infty\\ \label{eq:Asmpt2}
\frac{1}{1+ \left( r_{ij} /\eta\right)^{\upsilon}} \rightarrow \Phi( r_{ij};r_c) &  {\rm as } &\upsilon\rightarrow\infty,
\end{eqnarray}
where $r_c=\eta$ and $\Phi( r_{ij};r_c) $ is  the ideal low-pass  filter (ILF) used in the GNM Kirchhoff matrix
\begin{eqnarray}\label{eqn:IdealLF}
\Phi( r_{ij};r_c)  = \begin{cases}\begin{array}{ll}
       1, &  r_{ij} \leq r_c \\
       0,  & r_{ij}  > r_c  \\
			\end{array}
       \end{cases}.
\end{eqnarray}
Relations (\ref{eq:Asmpt1}) and (\ref{eq:Asmpt2}) unequivocally connect FRI correlation functions to the GNM Kirchhoff matrix. It is important to examine whether the ILF is still a FRI correlation function.  Mathematically, the ILF is a special real-valued monotonically decreasing correlation function and also satisfies admissibility conditions  (\ref{eq:couple_matrix1-1}) and (\ref{eq:couple_matrix1-2}). In fact,  all FRI correlation functions are  low-pass filters as well. Therefore, both   GNM and   FRI admit low-pass filters in their constructions. Indeed, GNM is very special in the sense that there is only one unique ILF, while, there are infinitely many other low-pass filters. Figure \ref{functions} illustrates the behavior and relation of the above low-pass filters or  correlation functions. Clearly, the ILF is completely localized for any given cutoff value. In general,   generalized exponential function and generalized Lorentz function are delocalized and the former decays faster than the latter for a given power. The combination of a low power value and a large scale gives rise to nonlocal correlations.  Our earlier test indicates that $\upsilon=3$ and $\eta=3$\AA~ provides a good flexibility analysis for   a set of 365 proteins \cite{Opron:2014}.  

%assess the protein network packing rigidity or density in dramatically different manners from   typical FRI correlation functions.

To further   bring to light the mathematical foundation of the GNM and FRI methods,  we  consider a generalized Kirchhoff matrix  \cite{KLXia:2013f, KLXia:2013b}
\begin{eqnarray}\label{eqn:MKirchhoff}
\Gamma_{ij}(\Phi )  = \begin{cases}\begin{array}{ll}
       - \Phi( r_{ij};\eta), &i\neq j  \\
        -\sum_{j, j\neq i}^N\Gamma_{ij}(\Phi),  & i=j
							\end{array}
       \end{cases},
\end{eqnarray}
where $\Phi( r_{ij};\eta)$ is an admissible FRI correlation function.  The generalized Kirchhoff matrix includes the Kirchhoff matrix as a special  case. It is important to note  that  each diagonal element is a  FRI rigidity index: $\mu_i=\Gamma_{ii}(\Phi )$. Therefore,  the generalized Kirchhoff matrix   provides a unified starting point for both the FRI and GNM methods. However, the striking difference between the GNM and FRI methods is that  to predict B-factors, the GNM seeks a matrix inverse of the Kirchhoff matrix (\ref{eqn:Kirchhoff}), whereas, the FRI takes the direct inverse of the diagonal   elements of the generalized  Kirchhoff matrix (\ref{eqn:MKirchhoff}).

Based on the above analysis, it is straightforward to construct   correlation function based GNMs via  the matrix inverse of the generalized Kirchhoff matrix (\ref{eqn:MKirchhoff}), which leads to infinitely many new GNMs, including the original GNM as a special limiting case. It is also possible to contruct to construct the FRI by using the Kirchhoff matrix, which gives rise to a unique FRI. Question arise as what are the relative performance of these correlation function based GNM and FRI methods. Another question is  whether there is any further relation between these  two distinguished approaches. Specifically, what is the relation between the diagonal  elements of the GNM matrix inverse  and the FRI direct inverse of the diagonal  elements, for a given generalized Kirchhoff matrix? To answer these questions, we select two representative correlation functions, i.e., the Lorentz ($\upsilon=3$) and ILF functions to construct the generalized Kirchhoff matrix (\ref{eqn:MKirchhoff}). The Lorentz function is a typical example for many   correlation functions studied in our earlier work  \cite{Opron:2014}. In contrast, the ILF function is an extreme case of FRI correlation functions. The resulting two generalized Kirchhoff matrices (\ref{eqn:MKirchhoff}) can be used for calculating the GNM matrix inverse or the inverse diagonal elements of the FRI matrix. This results in possible combinations or methods, namely, FRI-Lorentz, FRI-ILF, GNM-Lorentz and GNM-ILF. Performances of these methods are carefully analyzed. 

\begin{figure}[]
\begin{center}
\begin{tabular}{c}
\includegraphics[width=0.49\textwidth]{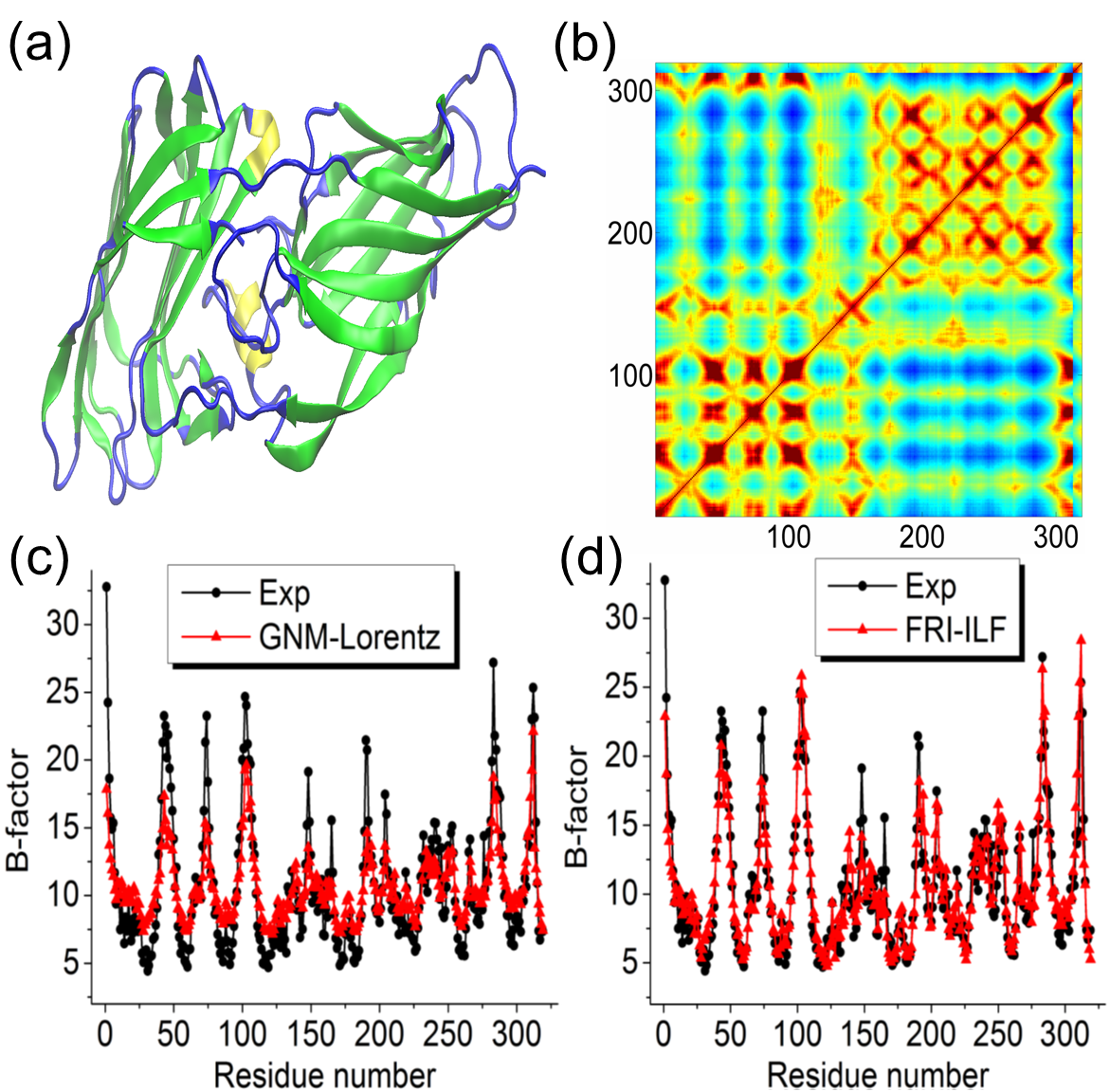}
\end{tabular}
\end{center}
\caption{Illustration of protein 2Y7L.
(a) Structure of protein 2Y7L having two domains;
(b) Correlation map generated by using GNM-Lorentz indicating two   domains;
(c) Comparison of experimental B-factors and those predicted by GNM-Lorentz ($\eta=16$\AA);
(d) Comparison of experimental B-factors and those predicted by FRI-ILF ($r_c=24$\AA).   %%%%%0.92
 }
\label{2Y7L}
\end{figure}

To answer the above mentioned questions, we first employ a pathogenic fungus Candida albicans (Protein Data Bank ID: 2Y7L) with 319 residues as shown in Fig. \ref{2Y7L}(a) to explore the aforementioned four methods. We consider the coarse-grained  C$_{\alpha}$ representation of protein 2Y7L. We denote $B^{\rm GNM-ILF}$, $B^{\rm FRI-ILF}$, $B^{\rm GNM-Lorentz}$ and $B^{\rm FRI-Lorentz}$ respectively  the predicted B-factors of 
GNM-ILF, FRI-ILF, GNM-Lorentz and  FRI-Lorentz methods. The experimental B-factors from X-ray  diffraction,  $B^{\rm Exp}$, are employed for a comparison.  The Pearson product-moment correlation coefficient (PCC) is used to measure  the strength of the linear relationship or dependence between each two sets of B-factors. To evaluate the performance of four methods, we compute the PCCs between predicted B-factors and experimental B-factors. Since performance of these methods depends on their  parameters, i.e.,  the cutoff distance ($r_c$) in the ILF or the scale value ($\eta$) in the Lorentz function, the theoretical B-factors are computed over a wide range of $r_c$ and $\eta$ values.

\begin{figure}[]
\begin{center}
\begin{tabular}{c}
\includegraphics[width=0.49\textwidth]{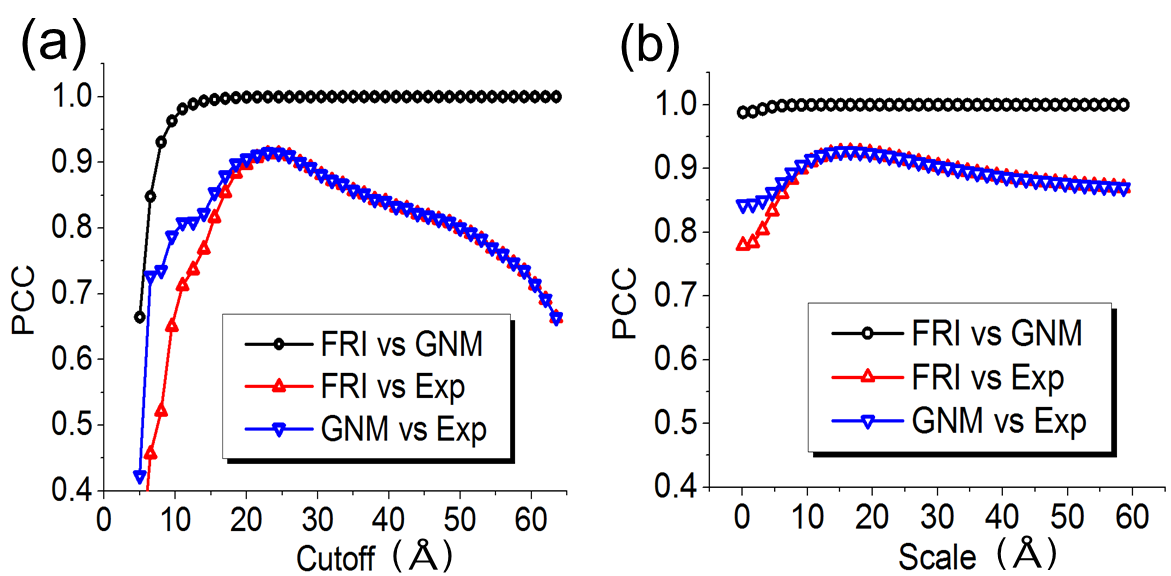}
\end{tabular}
\end{center}
\caption{PCCs  between various B-factors for protein 2Y7L.
(a) Correlations       between $B^{\rm GNM-ILF}$ and $B^{\rm Exp}$,
                       between $B^{\rm FRI-ILF}$ and $B^{\rm Exp}$, and
											 between $B^{\rm GNM-ILF}$ and $B^{\rm FRI-ILF}$;
(b) Correlations       between $B^{\rm GNM-Lorentz}$  and $B^{\rm Exp}$,
                       between $B^{\rm FRI-Lorentz}$  and $B^{\rm Exp}$, and
                       between $B^{\rm GNM-Lorentz}$ and $B^{\rm FRI-Lorentz}$.
}
\label{ccGNMFRI2Y7L}
\end{figure}

Figure \ref{ccGNMFRI2Y7L} depicts PCCs between various B-factors for protein 2Y7L. As shown in  Fig. \ref{ccGNMFRI2Y7L} (a),   the cutoff distance $r_c$ of the ILF is varied from 5\AA~ to 64\AA. The PCCs between $B^{\rm GNM-ILF}$ and $B^{\rm Exp}$, and between $B^{\rm FRI-ILF}$ and $B^{\rm Exp}$, indicate that both  GNM-ILF and    FRI-ILF are able to provide accurate predictions of the experimental B-factors. Their best predictions are attained around $r_c=24$\AA, which is significantly larger than the commonly used GNM cutoff distance of 7-9\AA, partially due to the fact that protein 2Y7L is relatively large.

It is interesting to observe that GNM-ILF and FRI-ILF provide essentially identical predictions when the cutoff distance is equal to or larger than 20\AA.  This phenomenon indicates that when the cutoff is sufficiently large, the diagonal elements of the GNM inverse matrix and the direct inverse of the diagonal elements of the FRI correlation matrix become linearly dependent.  To examine the  relation between GNM-ILF and FRI-ILF,  we compute PCCs  between   $B^{\rm GNM-ILF}$ and $B^{\rm FRI-ILF}$ over the same range of cutoff distances. As shown in  Fig. \ref{ccGNMFRI2Y7L}(a), there is a strong linear dependence between  $B^{\rm GNM-ILF}$ and $B^{\rm FRI-ILF}$ for $r_c\geq10$\AA. To understand this dependence at large cutoff distance, we analytically calculate $i$th diagonal element of the GNM  inverse matrix
\begin{eqnarray}\label{eqn:proprotion0}
\left(\Gamma^{-1}(\Phi( r_{ij};r_c\rightarrow\infty)) \right)_{ii} = \frac{N-1}{N^2}\rightarrow \frac{1}{N} ~{\rm as}~N\rightarrow\infty
\end{eqnarray}
and the FRI inverse of the $i$th diagonal element
\begin{eqnarray}  \label{eqn:proprotion01}
 \frac{1}{\sum_{j,j\neq i}^N  \Phi(r_{ij};r_c\rightarrow\infty)}=  \frac{1}{N-1} \rightarrow \frac{1}{N} ~{\rm as}~N\rightarrow\infty  .
\end{eqnarray}
These results elucidate the strong asymptotic correlation between $B^{\rm GNM-ILF}$ and $B^{\rm FRI-ILF}$ in Fig. \ref{ccGNMFRI2Y7L}(a). They also explain why predictions of the original GNM and FRI-ILF deteriorate as $r_c$ is sufficiently large because  all the predicted B-factors become identical, i.e., $\frac{N-1}{N^2}$ with $N=319$.  

% Note that the increase in the scale leads to a diagonally dominate matrix!!!

The performance and comparison between GNM-Lorentz and FRI-Lorentz are illustrated in Fig. \ref{ccGNMFRI2Y7L}(b) for the scale value $\eta$ from 0.5\AA~ to 64\AA. First, it is seen that the  GNM-Lorentz is a successful new approach. In fact, it outperforms the original GNM  for the peak PCCs. A comparison of the predicted B-factors and the experimental B-factors is plotted in Figs. \ref{2Y7L}(c) and \ref{2Y7L}(d) for  GNM-Lorentz and  FRI-ILF, respectively. It is seen that $B^{\rm FRI-ILF}$ more closely matches the experimental B-factors than $B^{\rm  GNM-Lorentz}$  does due to the different fitting schemes employed by two methods as shown in Eqs. (\ref{eqn:GNM}) and (\ref{eqn:FRI}), respectively.

As shown in Fig.  \ref{ccGNMFRI2Y7L}(b), the predictions from  GNM-Lorentz and FRI-Lorentz become identical as $\eta\geq5$\AA. A strong correlation between $B^{\rm GNM-Lorentz}$ and $B^{\rm FRI-Lorentz}$ is revealed at an even smaller  scale value.  This behavior leads us to speculate a general relation
\begin{eqnarray}\label{eqn:proprotion}
\left(\Gamma^{-1}(\Phi( r_{ij};\eta)) \right)_{ii}
 \longrightarrow
\frac{c}{\sum_{j,j\neq i}^N  \Phi(r_{ij};\eta)},  ~\eta \rightarrow \infty,
\end{eqnarray}
 where $c$ is a constant. Relation (\ref{eqn:proprotion}) means that the correlation function based GNM  is equivalent to the FRI for a given admissible correlation function when the scale parameter is sufficiently large. This relation is certainly true for the ILF as analytically proved in Eqs. (\ref{eqn:proprotion0}) and (\ref{eqn:proprotion01}).  Relation (\ref{eqn:proprotion})  is  a very interesting and powerful  result not only for sake of understanding GNM and FRI methods, but also for the design  of  accurate and efficient new methods.

\begin{figure}[]
\begin{center}
\begin{tabular}{c}
\includegraphics[width=0.49\textwidth]{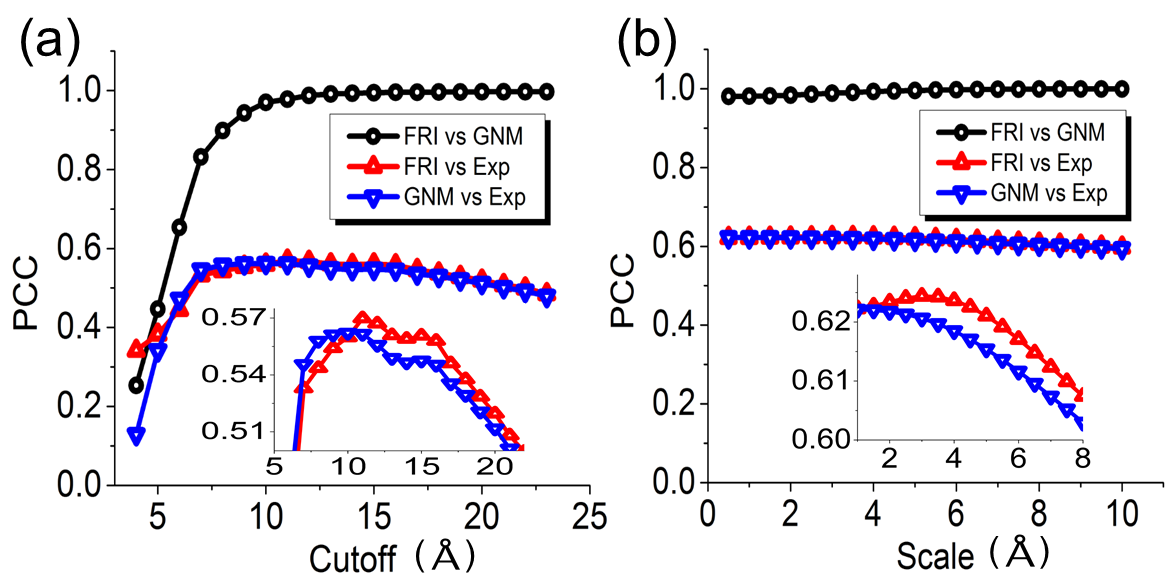}
\end{tabular}
\end{center}
\caption{ PCCs between various B-factors averaged over  364 proteins.
(a) Correlations       between $B^{\rm GNM-ILF}$ and $B^{\rm Exp}$,
                       between $B^{\rm FRI-ILF}$ and $B^{\rm Exp}$, and
											 between $B^{\rm GNM-ILF}$ and $B^{\rm FRI-ILF}$;
(b) Correlations       between $B^{\rm GNM-Lorentz}$  and $B^{\rm Exp}$,
                       between $B^{\rm FRI-Lorentz}$  and $B^{\rm Exp}$, and
                       between $B^{\rm GNM-Lorentz}$ and $B^{\rm FRI-Lorentz}$.
 }
\label{CC_F_G364}
\end{figure}

It remains to prove that the above findings from a single protein are translatable and verifiable to a large class of biomolecules. To this end, we consider a set of 364 proteins, which is a subset of the 365 proteins utilized and documented in our earlier work \cite{Opron:2014}. The omitted protein is 1AGN, which has been found to have unrealistic experimental B-factors.  We carry out systematic studies of four methods over a rang of cutoff distances or scale values. For each given $r_c$ or $\eta$, the PCCs between two sets of B-factors are averaged over 364 proteins. Figure \ref{CC_F_G364} illustrates our results. Figure \ref{CC_F_G364}(a) plots the results of the ILF   implemented in both GNM and FRI methods with the cutoff distance varied from 4\AA~ to 23\AA. Figure \ref{CC_F_G364}(b)  depicts  similar results obtained by using the Lorentz function implemented in two methods. The scale value is explored over the range of  0.5\AA~ to 10\AA.

First,  the proposed new method, GNM-Lorentz, is very successful for the B-factor prediction of 364 proteins as shown in Fig. \ref{CC_F_G364}(b). The best GNM-Lorentz  prediction  is about 10.7\% better than that  of the original GNM shown in Fig. \ref{CC_F_G364}(a). In fact, GNM-Lorentz outperforms the original GNM over a wide range of parameters for this set of proteins, which indicates that  the proposed generalization is practically valuable. Similarly, FRI-Lorentz  is also about 10\% more accurate than FRI-ILF in the B-factor prediction. Since the ILF is a special case and there are  infinitely many FRI correlation functions, there is a wide variety of correlation function based GNMs that are expected to deliver more accurate flexibility analysis than the original GNM does.

Additionally, the FRI-Lorentz method is able to attain the best average prediction for 364 proteins among four methods as shown in the zoomed in parts in Fig. \ref{CC_F_G364}(b).  However, for a given correlation function, the difference between FRI and GNM predictions is very small.

Moreover, for a given admissible FRI function, GNM and FRI B-factor predictions are strongly linearly correlated  and reach near 100\% correlation when $r_c>9$\AA~ or $\eta>0.5$\AA~ for 364 proteins as demonstrated in Fig. \ref{CC_F_G364}.  This finding offers a solid confirmation of  Eq. (\ref{eqn:proprotion}). Therefore, correlation function based GNMs, including the original GNM as a special case,  are indeed equivalent to the corresponding FRI methods in the flexibility analysis for a wide range of commonly used  scale values.

Furthermore, it has been shown that the fast FRI is a linear scaling method \cite{Opron:2014}, while GNM scales as $O(N^3)$ due to their matrix inverse procedure. As a result, the accumulated  CPU times for the B-factor predictions of  364 proteins at $r_c=7$ or $\eta=3$ are 0.88, 1.57, 5071.32  and 4934.79 seconds respectively for the FRI-ILF,  FRI-Lorentz, GNM-ILF and  GNM-Lorentz. In fact, GNM methods are very fast for small proteins as well.  Most of the accumulated GNM CPU times are due to the computation of three largest proteins (i.e., 1F8R, 1H6V and 1QKI) in the test set.

It is worth mentioning that  that the earlier FRI rigidity index includes the contribution from the self correlation, i.e., the diagonal term \cite{KLXia:2013d,Opron:2014}. The present findings do not change if  the summation in the generalized  Kirchhoff matrix (\ref{eqn:MKirchhoff}) is modified to include the diagonal term and then the calculation of GNM matrix inverse is modified to include the contribution from first eigen mode, i.e., $\left(\Gamma^{-1} \right)_{ii}=\sum_{k=1}^N  \lambda^{-1}\left[{\bf u}_k {\bf u}_k^T \right]_{ii}$.  In fact, this modification makes the   generalized  Kirchhoff matrix less singular and  fast converging.

%\section*{Acknowledgments}
 This work was supported in part by NSF grants   IIS-1302285,  and DMS-1160352, and NIH Grant R01GM-090208.
The authors  thank Jianyu  Chen and Michael Feig for useful  discussions and acknowledge the Mathematical Biosciences Institute for hosting valuable workshops.

\small
%\bibliographystyle{abbrv}
%\bibliographystyle{plain}
%\bibliographystyle{unsrt}
%\bibliography{refs}
%\bibliography{refs}

\begin{thebibliography}{10}

\bibitem{Alvarez-Garcia:2014}
Daniel Alvarez-Garcia and Xavier Barril.
%\newblock Relationship between protein flexibility and binding: Lessons for  structure-based drug design.
\newblock {\em Journal of Chemical Theory and Computation}, 10(6):2608--2614,
  2014.

\bibitem{Atilgan:2001}
A.~R. Atilgan, S.~R. Durrell, R.~L. Jernigan, M.~C. Demirel, O.~Keskin, and
  I.~Bahar.
%\newblock Anisotropy of fluctuation dynamics of proteins with an elastic  network model.
\newblock {\em Biophys. J.}, 80:505 -- 515, 2001.

\bibitem{Bahar:1998}
I.~Bahar, A.~R. Atilgan, M.~C. Demirel, and B.~Erman.
%\newblock Vibrational dynamics of proteins: Significance of slow and fast modes in relation to function and stability.
\newblock {\em Phys. Rev. Lett}, 80:2733 -- 2736, 1998.

\bibitem{Bahar:1997}
I.~Bahar, A.~R. Atilgan, and B.~Erman.
%\newblock Direct evaluation of thermal fluctuations in proteins using a single-parameter harmonic potential.
\newblock {\em Folding and Design}, 2:173 -- 181, 1997.

\bibitem{Brooks:1983}
B.~R. Brooks, R.~E. Bruccoleri, B.~D. Olafson, D.J. States, S.~Swaminathan, and
  M.~Karplus.
%\newblock Charmm: A program for macromolecular energy, minimization, and dynamics calculations.
\newblock {\em J. Comput. Chem.}, 4:187--217, 1983.

\bibitem{ZBu:2011}
Z.~Bu and D.~J. Callaway.
%\newblock Proteins move! protein dynamics and long-range allostery in cell signaling.
\newblock {\em Advances in Protein Chemistry and Structural Biology},
  83:163--221, 2011.

\bibitem{Dudko:2006}
O.~K. Dudko, G.~Hummer, and A.~Szabo.
%\newblock Intrinsic rates and activation free energies from single-molecule pulling experiments.
\newblock {\em Phys. Rev. Lett.}, 96:108101, 2006.

\bibitem{Fischer:2014}
Marcus Fischer, Ryan~G. Coleman, James~S. Fraser, and Brian~K. Shoichet.
%\newblock Incorporation of protein flexibility and conformational energy penalties in docking screens to improve ligand discovery.
\newblock {\em Nature Chemistry}, 6:575--583, 2014.

\bibitem{Flory:1976}
P.~J. Flory.
%\newblock Statistical thermodynamics of random networks.
\newblock {\em Proc. Roy. Soc. Lond. A,}, 351:351 -- 378, 1976.

\bibitem{Fraser:2009}
M.~W. Fraser, J. S.and~Clarkson, S.~C. Degnan, R.~Erion, D.~Kern, and T.~Alber.
%\newblock Hidden alternative structures of proline isomerase essential for catalysis.
\newblock {\em Nature}, 462:669--673, 2009.

\bibitem{Go:1983}
N.~Go, T.~Noguti, and T.~Nishikawa.
%\newblock Dynamics of a small globular protein in terms of low-frequency vibrational modes.
\newblock {\em Proc. Natl. Acad. Sci.}, 80:3696 -- 3700, 1983.

\bibitem{Halle:2002}
B.~Halle.
%\newblock Flexibility and packing in proteins.
\newblock {\em PNAS}, 99:1274--1279, 2002.

\bibitem{Hinsen:1998}
K.~Hinsen.
%\newblock Analysis of domain motions by approximate normal mode calculations.
\newblock {\em Proteins}, 33:417 -- 429, 1998.

\bibitem{Hinsen:2008}
K.~Hinsen.
%\newblock Structural flexibility in proteins: impact of the crystal environment.
\newblock {\em Bioinformatics}, 24:521 -- 528, 2008.

\bibitem{Jacobs:2001}
D.~J. Jacobs, A.~J. Rader, L.~A. Kuhn, and M.~F. Thorpe.
%\newblock {Protein flexibility predictions using graph theory}.
\newblock {\em {Proteins-Structure, Function, and Genetics}},
  {44}({2}):{150--165}, {AUG 1} {2001}.

\bibitem{Kondrashov:2007}
D.~A. Kondrashov, A.~W. Van~Wynsberghe, R.~M. Bannen, Q.~Cui, and Jr. G.~N.
  Phillips.
%\newblock Protein structural variation in computational models and crystallographic data.
\newblock {\em Structure}, 15:169 -- 177, 2007.

\bibitem{Levitt:1985}
M.~Levitt, C.~Sander, and P.~S. Stern.
%\newblock Protein normal-mode dynamics: Trypsin inhibitor, crambin, ribonuclease and lysozyme.
\newblock {\em J. Mol. Biol.}, 181(3):423 -- 447, 1985.

\bibitem{Marsh:2014}
J.~A. Marsh and S.~A. Teichmann.
%\newblock Protein flexibility facilitates quaternary structure assembly and evolution.
\newblock {\em PLoS Biol}, 12(5):e1001870, 2014.

\bibitem{Opron:2014}
K.~Opron, K.~L. Xia, and G.~W. Wei.
%\newblock Fast and anisotropic flexibility-rigidity index for protein flexibility and fluctuation analysis.
\newblock {\em Journal of Chemical Physics}, 140:234105, 2014.

\bibitem{JKPark:2013}
J.~K. Park, Robert Jernigan, and Zhijun Wu.
%\newblock Coarse grained normal mode analysis vs. refined gaussian network model for protein residue-level structural fluctuations.
\newblock {\em Bulletin of Mathematical Biology}, 75:124 --160, 2013.

\bibitem{Tasumi:1982}
M.~Tasumi, H.~Takenchi, S.~Ataka, A.~M. Dwidedi, and S.~Krimm.
%\newblock Normal vibrations of proteins: Glucagon.
\newblock {\em Biopolymers}, 21:711 -- 714, 1982.

\bibitem{Thorpe:2007}
M.~F. Thorpe.
%\newblock Comment on elastic network models and proteins.
\newblock {\em Physical Biology}, 4:60--63, 2007.

\bibitem{Tirion:1996}
M.~M. Tirion.
%\newblock Large amplitude elastic motions in proteins from a single-parameter,
  atomic analysis.
\newblock {\em Phys. Rev. Lett.}, 77:1905 -- 1908, 1996.

\bibitem{GWei:2000}
G.~W. Wei.
%\newblock Wavelets generated by using discrete singular convolution kernels.
\newblock {\em Journal of Physics A: Mathematical and General}, 33:8577 --
  8596, 2000.

\bibitem{KLXia:2013d}
K.~L. Xia, K.~Opron, and G.~W. Wei.
%\newblock Multiscale multiphysics and multidomain models --- { Flexibility} and rigidity.
\newblock {\em Journal of Chemical Physics}, 139:194109, 2013.

\bibitem{KLXia:2013f}
K.~L. Xia and G.~W. Wei.
%\newblock A stochastic model for protein flexibility analysis.
\newblock {\em Physical Review E}, 88:062709, 2013.

\bibitem{KLXia:2013b}
K.~L. Xia and G.~W. Wei.
%\newblock Three-dimensional {MIB} galerkin method for elliptic interface problems.
\newblock {\em submitted to Journal of Computational Physics}, 2013.

\bibitem{LWYang:2008}
L.~W. Yang and C.~P. Chng.
%\newblock Coarse-grained models reveal functional dynamics--{I}. elastic network models--theories, comparisons and perspectives.
\newblock {\em Bioinformatics and Biology Insights}, 2:25 -- 45, 2008.

\end{thebibliography}

\end{document}